\documentstyle[12pt]{article}

\newcommand{\be}{\begin{equation}}
\newcommand{\ee}{\end{equation}}

\catcode `\@=11
\@addtoreset{equation}{section}

\def\section{\@startsection {section}{1}{\z@}{-3.5ex plus -1ex minus
     -.2ex}{2.3ex plus .2ex}{\normalsize\bf}}
\def\subsection{\@startsection{subsection}{2}{\z@}{-3.25ex plus -1ex 
minus
 -.2ex}{1.5ex plus .2ex}{\normalsize\bf}}
 
\catcode `\@=12

\begin{document}

\title{Inhomogeneous imperfect fluid  \\
 spherical models without  \\
Big-Bang singularity} 
\author{Naresh Dadhich, \thanks{E-mail : nkd@iucaa.ernet.in} \\
Inter-University Centre for Astronomy \& Astrophysics,\\
Post Bag 4, Ganeshkhind, Pune - 411 007, India.}
\date{}
\maketitle

\begin{abstract}
So far all known singularity-free cosmological models are
cylindrically symmetric. Here we present a new family of spherically
symmetric non-singular models filled with imperfect fluid
and radial heat flow, and satisfying the weak and strong
energy conditions. For large $t$ anisotropy in pressure 
and heat flux tend
to vanish leading to a perfect fluid. There is a 
free function of time in the model, 
which can be suitably chosen
for non-singular behaviour and there exist multiplicity of 
such choices. \\
\end{abstract}


\vfill
\begin{flushright} IUCAA-40/97 \end{flushright}
\vfill

\newpage

\section{Introduction}

\noindent Although the present day observations indicate that the Universe 
at  large  scale  is homogeneous and isotropic  and  it  is  well 
described by the standard Friedman-Robertson-Walker (FRW)  model. 
It  is  also  a  well  recognised  view  that  consideration   of 
inhomogeneity at early times is well in order for generic initial 
conditions   and  for  facilitating  formation  of  large   scale 
structures in the Universe. Further it is now known for some time 
[1-4]  that  inhomogeneous spacetime admits a family  of  perfect 
fluid cosmological models without the big- bang singularity. Note 
that  these models are exact solutions of the  Einstein  equation 
satisfying  the causality and energy conditions, yet there is  no 
divergence  of any physical and geometric  parameters  throughout 
the  spacetime.  This happens because spacetime  does  not  admit 
compact  trapped  surfaces which invalidates application  of  the 
singularity theorems [5]. \\

\noindent From the Raychaudhuri equation [6], it is clear that singularity 
can  be avoided only if acceleration or rotation is non-zero.  In 
cosmology  rotation is generally not  favoured  and  hence 
acceleration  must  be present if singularity is to  be  avoided. 
This  means spacetime has to be inhomogeneous. In all  the  known 
non-singular  solutions,  shear  is  also  non-zero,   indicating 
anisotropy  as  well. It can be proved for a  general  orthogonal 
metric separable in space and time variables [4-5] and for a  G2-
symmetric perfect fluid model that presence of shear is essential 
for  presence  of  acceleration  [7].  Though  shear  contributes 
positively  to the effective gravitational charge density in  the 
Raychaudhuri  equation,  its dynamical role  as  making  collapse 
incoherent  can  combine  well  with  acceleration  in   avoiding 
singularity.  Thus shear also seems to be playing very  important 
role in non-singular models. \\
        
\noindent So  far all known non-singular solutions are  inhomogeneous  and 
isotropic as well as have cylindrical symmetry [1-4,8-9]. It  has 
perhaps their cylindrical symmetry that comes in the way of their 
application  in practical cosmology. Since the Universe is  known 
to be spherical to a good degree, it is pertinent to ask  whether 
it  is  possible  to have a  spherically  symmetric  cosmological 
model?  In this note we give a singularity-free prescription  for 
imperfect  fluid  with heat flux. This is a  prescription  rather 
than  a  solution  for no equation has  been  solved.  A  general 
spherically  symmetric metric has only four independent stresses,  which 
could be interpreted as density, radial and transverse pressures, 
and  radial  heat  flux. This will be true  for  any  spherically 
symmetric metric. The question is to give a prescription which is 
free of singularity and has proper desired behaviour for physical 
parameters. This has been achieved by letting the Tikekar  static 
model  [10],  which is a particular case of the  Tolman  solution 
[11], expand. \\

\noindent Our  model  represents a spherically  symmetric  cosmological 
universe  filled  with  imperfect fluid having  unequal  radial  and 
transverse pressure and radial heat flow. The spacetime satisfies 
the weak and strong energy conditions. It 
is free of any kind of singularity as all the physical as well as 
kinematic  parameters  remain finite and regular  in  the  entire 
range of the variables $t$ and $r$. It has the typical behaviour of a 
non-singular  model; density and pressure vanishing for  large  $t$ 
and $r $ and being maximum at $t = 0$ and $r = 0,$ expansion  parameter 
and  radial heat flux change their sense at $t = 0$, while acceleration
tends to zero as $r \longrightarrow 0$. There is however an unusual
feature that heat flows radially inward as the universe expands.
The heat flow vanishes at $r$ or $ t \longrightarrow 0$ as well
as $r$ or $\pm t \longrightarrow \infty$ and it falls off as $t^{-4}$
or $r^{-4}$ asymptotically. The pressure anisotropy vanishes for
$r \longrightarrow 0$ as well as for $r$ or $\pm t \longrightarrow
\infty$. That is, asymptotically it tends to perfect fluid.
The  shear  is 
also  non-zero  and  hence the model is  both  inhomogeneous  and 
anisotropic.  \\

\section{The model}

\noindent The model is described by the spherically symmetric metric,

\begin{equation}
ds^2 = (r^2 + P) dt^2 - \frac{2r^2 + P}{r^2 + P} dr^2 - r^2
(d \theta^2 + \sin^2 \theta d \varphi^2)
\end{equation}

\noindent where $P = P(t)$. The  imperfect 
fluid is represented by the energy-momentum tensor [12], 

\begin{equation}
T_{ik} = (\rho + p) u_i u_k - p g_{ik} + \bigtriangleup p [c_i c_k
+ \frac{1}{3} (g_{ik} - u_i u_k)] + 2 q c_{(i} u_{k)} 
\end{equation}

\noindent where $u_i$ and $c_i$ are respectively unit timelike and
spacelike vectors, $\rho$  the energy density,
$p$ the isotropic fluid pressure, $\bigtriangleup p$
the pressure anisotropy and $q$  heat flux. \\  

\noindent We employ the comoving coordinates to write $u_i = \sqrt{g_{00}}
\delta^0_i$ and take $c_i = \sqrt{g_{11}} \delta^1_i$.
The kinematic parameters; expansion, shear and 
acceleration for the metric (1) read as follows: 

\begin{equation}
\theta = \frac{- \dot P r^2}{2(2r^2 + P) (r^2+ P)^{3/2}}, ~
\sigma^2 = \frac{2}{3} \theta^2, ~ \dot u_r = -\frac{r}
{r^2 + P}.
\end{equation}

\noindent Now applying the Einstein equation, we obtain \\

\begin{equation}
8 \pi \rho = \frac{2 r^2 + 3P}{(2r^2 + P)^2} 
\end{equation}

\begin{equation}
8 \pi p_r = \frac{1}{2r^2 + P} 
\end{equation}

\begin{equation}
8 \pi p_{\perp} = \frac{1}{2r^2 + P} + \frac{r^2}{4(2 r^2 + P)(r^2 + P)^2}
\left[2 \ddot P - \frac{(9r^2 + 5P) \dot P^2}{(2r^2 + P)(r^2 + P)} \right]
\end{equation}

\begin{equation}
8 \pi q = \frac{- \dot P r}{(2 r^2 + P)^{3/2}(r^2 + P)}.
\end{equation}

\noindent The pressure anisotropy $\bigtriangleup p = p_r - p_{\perp}$ is given by

\begin{equation}
8 \pi \bigtriangleup p = \frac{- r^2}{4(2 r^2 + P)(r^2 + P)^2}
\left[2 \ddot P - \frac{(9r^2 + 5P) \dot P^2}{(2r^2 + P)(r^2 + P)} \right].
\end{equation}

\noindent Now we just need to make a suitable choice for $P(t)$ such that
all physical and kinematic parameters remain finite and regular.
Any even function of $t$ without zero would be an appropriate choice, 
for instance $P = \alpha^2 + \beta^2 t^2$ or $\cosh^a kt$ and so on. Thus 
there exists a family of non-singular models. In general $P(t)$
is free similar to the scale factor in FRW models. In here there is
a great deal of freedom and one may as well ask that any spherically
symmetric metric could be viewed as representing an imperfect fluid with radial heat flux. This is however true. 
Even then it is a non-trivial matter to have acceptable behaviour
for model satisfying energy conditions and being free of singularity. 
As far as we know this is for the first time such a proposal
has been made.\\

\noindent Let us consider the simplest case, $P = \alpha^2 + \beta^2 t^2$. 
It is easy to verify that all  the  above 
physical parameters as well  as  the  kinematic 
parameters remain finite and regular for the  entire 
range  of  variables; $- \infty < t < \infty,~ 0 \leq r < \infty$.
This  indicates  clearly 
that  the model is free of singularity. All the above  parameters 
tend  to  zero for $t \longrightarrow \pm \infty$
and/or $r \longrightarrow \infty$. Asymptotically it is  low 
density  universe,  which when contracts,  attains  the  maximum 
density at $t =0$ and $r = 0$, specified by the parameter 
$\alpha (\rho_{max} = 3/8 \pi\alpha^2),$ which can be chosen as small
as one pleases to have as large $\rho_{max}$ as 
desired. For large $t, \beta$ will correspond to the Hubble
parameter. At $t = 0$, the expansion parameter changes sign (contraction
$\longleftrightarrow $ expansion) and the acceleration tends to
zero as $r \longrightarrow 0$. The universe starts from low
density, contracts to high density and again expands to low
density state without encountering singular behaviour of any kind.
Anisotropy in pressure $\bigtriangleup p$ tends to zero for $r 
\longrightarrow 0$ as well as asymptotically ($t \longrightarrow
\pm \infty $ or $r \longrightarrow \infty$). For small $r$ and
large $t$, it 
approximates to the radiation universe, $\rho = 3p$, while
for large $r$ it tends to an isothermal stiff fluid with
$\rho = p \sim 1/r^2$ [13].
 $\rho $ and $p_r$ fall off as $t^{-2}$,
whereas $\bigtriangleup p$ falls off as $t^{-6}$ and $q$
as $t^{-4}$, which indicate
that fluid turns almost perfect (with isotropic pressure) for
large enough $t$. \\

\noindent We shall now verify that the model satisfies the weak and
strong energy conditions;
$T_{ik} \xi^i \xi^k \geq 0$ and  $(T_{ik} - \frac{1}{2}
T g_{ik}) \xi^i \xi^k \geq 0$ for any non-spacelike vector $\xi^i$.
We have

$$ T_{ik} \xi^i \xi^k = n^2(\rho + p_r - \bigtriangleup p)
-m^2 (p_r - \bigtriangleup p) $$

\noindent and

$$(T_{ik} - \frac{1}{2} T g_{ik}) \xi^i \xi^k = \frac{1}{2}
\left[(2 n^2 - m^2)(\rho + p_r - \bigtriangleup p)
+ m^2 (2 p_r - \bigtriangleup p) \right] $$

\noindent where $\xi_i u^i = n, \xi_i \xi^i = m^2$
and $n^2 \geq m^2$ always. It is clear from eqns (4), (5)
and (8) that the weak and strong energy conditions will be
satisfied for $P(t) = \alpha^2 + \beta^2 t^2$. A lengthy but
straight forward calculation shows that the model cannot
however satisfy the dominant energy condition (the vector
$T_{ik} \xi^k$ is non-spacelike). \\

\noindent All this is very fine but there is a discomforting
feature indicated
by $q \theta < 0$ (from (3) and (7)) in general independent of
specific choice for $P(t)$. This implies that there is a radially
inward heat flux for expanding phase and outward for contracting
phase. It may be noted that for the particular model under consideration,
$q=0$ for $r$ or $t=0$ and it falls off as $r^{-4}$ or $t^{-4}$
for large
$r$ or $t$. However, we must confess that this is rather
an unusual feature. \\

\noindent The  overall  evolution  of the model is  typical  of non-
singular models [3,4]; asymptotically low density passing through 
the  dense  state  at t = 0,  where  interchange  occurs  between 
expansion  and  contraction on one hand and  between  inflow  and 
outflow  for  the radial heat flux on the other.  The  remarkable 
features  of  this  model are: (a) it is  free  of  big-bang 
as well as any other singularity, (b) it is spherically symmetric,
which augurs well 
with  the symmetry of the realistic Universe, (c) for small $r$ 
and large $t$ it 
approximates to the radiation Universe, $\rho = 3p$, 
(d) asymptotically $(t \longrightarrow \pm \infty $
and/or $r \longrightarrow \infty$), the pressure anisotropy
and heat flux vanish leading to perfect fluid with $\rho = 
3p \sim 1/t^2$ for large $t$ and $\rho = p \sim 1/r^2$
for large $r$,
(e) the parameter $\alpha $ defines the maximum density
$(\rho_{max} = 3/8 \pi \alpha^2$ for $t = 0, r = 0$)
and $\beta$ will correspond to the Hubble parameter for large $t$,
(f) the
expansion parameter $\theta$ and heat flux $q$
change sign at $t = 0$ and fall off to zero as
$t \longrightarrow \pm \infty$ and/or $r \longrightarrow \infty$
and $q \theta < 0$,
(g) the acceleration $\dot u_r$ tends to zero as $r \longrightarrow
0$ and also falls off to zero asymptotically ($t \longrightarrow
\pm \infty$ and/or $r \longrightarrow \infty$), and (h) it is
evidently causally stable and obeys the weak 
and strong energy conditions. It however does not
satisfy the dominant energy condition,
which may perhaps be responsible for unusual behaviour for heat flow
(radially inward as the universe expands). It has also been argued
that at high densities, the possibility of violation of the
dominant energy condition
cannot be ruled out [14]. That means at high density $dp/d\rho$
may not always measure the true signal propogation
velocity and hence even if $dp/d\rho \geq 1$ there
is no conflict with special relativity. Thus possibility
of $p \geq \rho$ cannot be ruled out. In the context of our 
model, it suggests that at high density some unusual
behaviour may be permissible. However the model is free
of singularity and satisfies the weak and strong energy conditions. \\

\noindent One of the main objections against the so far known non-singular 
cosmological solutions was that they did not accord to  spherical 
symmetry  and hence their application to realistic cosmology  was 
greatly marred. By finding a family of spherically symmetric non-
singular  models we have overcome  this  objection  quite 
successfully and hence paving way for their practical application
in  cosmology. It should be noted that the model is as general as
FRW with $P(t)$ being free. Non-singular character will however
constrain the choice for $P(t)$, but even then there would remain
good deal of freedom. It would therefore be pertinent to examine its
cosmological viability. This is what we would like to do next.
Finally it is quite probable that there could similarly
exist other families
of non-singular spherical models which would make it all
very exciting. \\

\noindent {\bf Acknowledgement :} It is a pleasure to thank Roy Maartens
and Sunil Maharaj for helpful discussion through e-mail.

\end{document}